\documentclass[twocolumn,aps,prl]{revtex4}

\usepackage{graphicx}
\usepackage{float}
\usepackage{dcolumn}
\usepackage{amsmath}
\usepackage{amssymb}

\input{epsf}

\newcommand{\ie}{{\em i.e. }}

\def\be{\begin{equation}}
\def\ee{\end{equation}}
\def\bea{\begin{eqnarray}}
\def\eea{\end{eqnarray}}

\def\ie{{\it i.e.,\ }}

\def\half{{1 \over 2}}
\def\quarter{{1 \over 4}}

\begin{document}

\title{
\noindent\hfill\hbox to 1.5in{\rm  } \vskip 1pt \noindent\hfill\hbox
to 1.5in{\rm SLAC-PUB-14350 \hfill  } \vskip 1pt
\vskip 10pt
A Different Look at Dark Energy and the Time Variation of Fundamental Constants\footnote{This work was
supported by the U.~S.~DOE, Contract No.~DE-AC02-76SF00515.}}
\author{Marvin Weinstein$^1$\\
\normalsize{$^{1}$SLAC National Accelerator Laboratory, Stanford, CA, USA}\\
}

\begin{abstract}
This paper makes the simple observation that a fundamental length, or cutoff, in
the context  of Friedmann-Lema\^itre-Robertson-Walker (FRW) cosmology implies very different
things than for a static universe.  It is argued that it is reasonable to assume that
this cutoff is implemented by fixing the number of quantum degrees of freedom per co-moving
volume (as opposed to a Planck volume) and the relationship of the
vacuum-energy of all of the fields in the theory to the cosmological constant (or dark energy)
is re-examined. The restrictions that need to be satisfied by a generic theory to
avoid conflicts with current experiments are discussed, and it is shown that in any theory
satisfying these constraints knowing the difference between $w$ and minus one allows one to
predict $\dot{w}$. It is argued that this is a robust result and if this prediction fails the idea of a fundamental cutoff of the type being discussed can be ruled out.
Finally, it is observed that, within the context of a specific theory, a co-moving
cutoff implies a predictable time variation of fundamental constants.  This is accompanied
by a general discussion of why this is so, what are the strongest phenomenological limits upon
this predicted variation, and which limits are in tension with the idea of a
co-moving cutoff.  It is pointed out, however, that a careful comparison of the predicted
time variation of fundamental constants is not possible without restricting to a particular
model field-theory and that is not done in this paper.
\end{abstract}
\maketitle

\section{Introduction}

In conventional field theory a cutoff is introduced as a trick to enable us to compute.  Eventually renormalization removes the cutoff. This formalism is developed for a universe that is static in time (\ie Minkowski space), so one doesn't spend a great deal of time thinking about how the notion of a cutoff works in a universe that is expanding.  This paper presents an argument that there are virtues to taking the idea of a cutoff seriously and asking what that means in the context of an expanding universe.

In a static universe a fixed cutoff implies a finite number of quantum mechanical degrees of freedom per Planck volume. If the universe is expanding there are two possibilities one can entertain for how the cutoff behaves as the universe grows larger.  The first is that the number of degrees of freedom per Planck volume remains fixed and more degrees of freedom are created to realize this assumption;  the second, the one that I favor, is that the number of degrees of freedom per co-moving volume remains fixed and perforce the number of degrees of freedom per Planck volume goes down.  My goal in this paper is to argue that this point of view leads to a very different way of looking at the question of the ``cosmological constant''.  This general argument has the virtue that if $\dot{w}$ is measured to be larger than the value predicted from our knowledge of $w$, then the idea will be inescapably shown to be false.

It is not my aim to present a ``theory of everything'' or even a model that realizes the physics I discuss.  Instead I want to focus on the phenomenological implications of the general idea.  In particular, I want to ask what constraints any model of short distance physics will have to satisfy in order to keep this notion out of conflict with current experiments and what it will take to kill the idea.  The result of this discussion will be that, with one exception, surprisingly few constraints have to be satisfied in order to make things work.  The potential payoff is that one arrives at a picture of the universe wherein there is a very intimate connection between cosmology and particle physics, far more so than in our usual way of looking at things.  In particular, I will argue that if this scenario is correct, then it may be possibile to look at the ``Why now?'' question in a way that has nothing to do with anthropic arguments.  Rather, one will look for theories where the changing scale factor of the universe can drive phase transitions, so that ``Why now?'' becomes ``For what theories does physics look the way it looks now when the scale factor has the value it has now?''.

The first question to address is ``Why should the fundamental length vary with the scale factor of the universe?''.  I would argue that this idea is forced upon us by another problem that arises when one talks about cutoffs in a universe whose size is changing; namely, how one counts degrees of freedom in a cutoff theory. Imagine, for example, that we are working in a box of finite volume, $L^3$, and assume periodic boundary conditions.  In this case, in momentum space, momenta are discrete and take the values
\be
    k_x, k_y, k_z = {2 \pi n_x\over L}, {2 \pi n_y \over L}, {2 \pi n_z\over L}
\ee
where $L$ stands for the any one linear dimension of the box.  The
integers $n_x, n_y, n_z$ run from minus infinity to infinity.
In a finite volume there are no infra-red divergences of the theory due to discrete
nature of the momenta, but the ultra-violet divergences of the theory are there because
the ranges of the $n$'s are unbounded.  Putting in a finite momentum cutoff (or finite
energy cutoff) amounts to specifying a number $\Lambda = 2 \pi N_{max}/ L $
and requiring that  the absolute values of the components of the momentum must be
less than $\Lambda$.   If, as is the case in a field theory, we say that for each field
there is one  quantum mechanical degree of freedom associated to each $k$ value,
then the total number of degrees of freedom for this field is of order $(2 N_{max})^3$.
Now if we think of $a = 1/\Lambda$ as the fundamental length corresponding to
the momentum cutoff, then there are $\Lambda L / \pi  = L /\pi a \sim 2 N_{max}$
fundamental lengths along each dimension or $(L / \pi a)^3 \sim (2 N_{max})^3$
fundamental volumes in the universe.  It is conventional to say that we have one
quantum mechanical degree of freedom per field per fundamental volume.  If we take
$\Lambda$ to be the Planck scale, $M_p$, then we have one degree of freedom per
Planck volume.  The question is, ``what happens to these notions as the universe
expands?''.

Clearly, if after some finite time the universe has doubled in size, then
we have twice as many fundamental volumes as we had before.  Does this mean
that we have twice as many degrees of freedom?  In essence, if we do have twice as
many degrees of freedom, then we have to invent a mechanism for creating degrees of
freedom as the universe grows in size.  People have worked on this idea without
notable success and yet conventional field theory seems to accomplish exactly this
feat without difficulty.  How does this work?

Briefly stated, conventional field theory works by stealing degrees of freedom from the region above the cutoff (\ie the extra degrees of freedom come from the ultra-violet divergences of the theory). Simply put, we already noted that momenta are discretized in units of $2 \pi /L$ and the maximum momentum that lies below a cutoff $\Lambda$ is $N_{max} = \Lambda L/ \pi$. Thus, as the universe grows and $L$ gets bigger, momenta become more closely spaced and $N_{max}$ grows (\ie more momenta fit below the cutoff).  If we take seriously that at some point in the past there were only a finite number of degrees of freedom per fundamental volume and there is no reservoir of degrees of freedom available for creating new degrees of freedom as the universe expands, then we can only conclude that as the universe expands $N_{max}$ stays fixed and the effective cutoff comes down.  Another way of stating this idea is to say that in an expanding universe it is not the number of degrees of freedom per Planck volume that stays fixed, rather it is the number of degrees of freedom per {\it co-moving volume\/}.  This, of course, is exactly how things would work if there was a lattice at some very short distance scale and if the lattice spacing was coupled to the change in scale factor of the universe. So, what are the implications of having a fixed, finite number of degrees of freedom per co-moving volume?

First and foremost one has to observe that the total vacuum energy density coming from  all of the quantum degrees of freedom in the theory cannot, as is usually supposed, be a ``cosmological constant'', since it cannot be constant.  The mere fact that the number of degrees of freedom per unit volume decreases with time tells us that the expansion of the universe reduces the ground state energy contribution of the quantum degrees of freedom to the vacuum energy density.  The thing we need to understand in some detail is how rapidly this happens, and what the experimental consequences of this diminution of the vacuum energy density will be.  In the next section I will show how this works for the case of free fields, since the story is, to me, somewhat surprising.  In particular, with mild assumptions about the degrees of freedom in the theory we find that one can get a time-dependent ``cosmological constant'', or  time dependent ``dark energy density''.  We will see that with reasonable assumptions  this dark energy density contribution has a value for the parameter $w$, the parameter relating the pressure due to this vacuum contribution to the energy density, that is close to minus one.  One  also finds that, given the assumption that the number of degrees of freedom per co-moving  volume is constant, there is a relationship between the difference of $w$ from $-1$ and the value of $\dot{w}$.

The second issue that springs to mind arises when one moves beyond non-interacting fields and asks what are the implications for this idea when interactions are taken into account. Here the discussion is perforce murkier, although I will try to provide an argument for how, again with slightly tighter physical assumptions, the general story can remain unchanged. However, we will see that even if the ``dark energy density'' story remains the same, we now must to worry about the way in which coupling constants and masses change with time.  We will see that the strongest bounds on any theory of this type come from the limits on the change in the fine structure constant and the electron mass since the time of big-bang nucleosynthesis(BBN), and from limits on the change of $\alpha_{em}$ as seen by experiments looking at the Lyman-$\alpha$ forest. Surprisingly we will see that the one additional constraint on the theory needed to hide changes in $\alpha_{em}$ from the Lyman-$\alpha$ forest measurements suffices to keep the change in $\alpha_{em}$ from the time of BBN to now unobservable (at least at present).  One might worry that, the most serious problem for this idea comes from geological constraints on the time variation of $\alpha_{em}$, such as the Oklo phenomenon, since the bound is so strong.  Evading these limits would seem to require some ugly machinations; however, that is only true if one uses computations based upon a uniform FRW cosmology.  If, as I will point out later, one realizes that a region containing gravitationally bound matter has decoupled from the expansion of the universe, then in that region the co-moving volume is no longer expanding and thus, in that region, coupling constants aren't running.  Thus, since the Oklo reactor is an earth-bound reactor, we don't expect $\alpha_{em}$ to have changed over the history of the earth and so the Oklo bound is irrelevant.

\section{Preliminaries}

Let us begin with by assuming that we are talking in the context
of FRW\cite{FRW} cosmology, where the line element is of the form
\be
   ds^2 = dt^2 - a(t)^2 d\vec{x}^2 .
\ee
(To save space I will simply write $a$ for $a(t)$ in what follows.)
Since we will be discussing a theory with some sort of fixed fundamental
length I will assume that the coordinates have dimensions and the scale factor, $a$,
is dimensionless. Furthermore, I assume that everything is in a preferred frame; \ie the
rest frame of the CMB radiation.
For the sake of simplicity let us begin by discussing a single
real scalar field $\phi(x)$.  In FRW cosmology the Lagrangian for this system
is
\bea
{\cal L}(x) &=& \sqrt{a^6} \,\half\,\left({d \phi(x) \over dt} \right)^2\nonumber \\
 &-& \half\,\sqrt{a^6}\, \left({1 \over a^2} (\nabla\phi(x))^2
 + m^2 \phi(x)^2\right) ,
\eea
where the factor of $\sqrt{a^6}$ multiplying everything is the square root of the
determinant of the metric that one has to have to define the volume integral.
To form the Hamiltonian we observe that the momentum conjugate to $\phi(x)$ is
$ \Pi(x) = a^3\,{d \phi(x) / dt}$ and so the Hamiltonian density becomes
\be
   H(x) = {\Pi(x)^2 \over 2\,a^3} + {a^3 \over 2} \,\left(  \left({\nabla \phi(x) \over a}\right)^2 + m^2 \,\phi(x)^2
   \right) .
\label{ffhamiltonian}
\ee
The key point to observe is that the factor of $a^3$ in this Hamiltonian is like the factor
of mass in the simple harmonic oscillator
\be
    H_{osc} = { p^2 \over 2\,m} + \half m\,\omega^2\,x^2 ,
\ee
it enters into the width of the ground state wave function, but not the energy.  The
energy is $ \omega / 2$.  Transforming the Hamiltonian, Eq.\ref{ffhamiltonian},
into momentum space we see that the vacuum energy density for a single real free scalar field,
in a theory with a maximum momentum cutoff for any value of $a$, is given by
\be
    {\cal E} = {1 \over 2 a^3} \int_0^\Lambda d^3 k \sqrt{ \left(k \over a\right)^2 + m^2} ,
\ee
where the factor of $a^{-3}$ appears because momenta have been defined for $a = 1$.
The formula for a complex scalar field will be the same, except that it is multiplied
by a factor of two.
Expanding the resulting integral for a complex field in powers of $a$ gives the following result:
\be
   {\cal E} \sim \quarter\,{\Lambda^4 \over a^4} + \quarter\,{\Lambda^2 m^2 \over a^2}
    + {1 \over 32} \,m^4 + {1 \over 8}\,m^4\,\ln\left(m\,a \over 2\,\Lambda\right)
    +  \ldots  .
\label{vacenergydensity}
\ee
The terms not shown are a series in powers of $ (a^2\,m^2 / \Lambda^2)^p m^4$,
and are negligible for large $\Lambda$.
It is equally easy to show that a similar formula is true for a fermion, except that
the energy density of the lowest eigenstate of the theory will have a minus sign in
front of all of the terms and each spin state will contribute like a complex scalar
field.

There are several things worth noting about the formula in Eq.\ref{vacenergydensity}.
First, the term quartic in
${\Lambda / a}$ is independent of mass and falls off like $a^{-4}$.
Hence, this term cannot be a cosmological constant, rather it is a contribution to the energy density that behaves like radiation.  Even if this term appeared at full strength, it would only represent a very large correction to the radiation energy density of the universe, not a factor that causes exponential growth of the scale factor.  If in addition we have
equal numbers of bosonic and fermionic degrees of freedom (counting degrees of freedom
as in super-symmetry) the individual contributions cancel exactly.

Similarly, the term quadratic in $\Lambda / a$ behaves like curvature.  It too can't be a cosmological constant.  Nevertheless it will generically be too large, given current bounds on the curvature contribution to FRW cosmology.  Fortunately, we can eliminate, or make it sufficiently small, by imposing the constraint that the sum of the boson masses squared minus the sum of the fermion masses squared is extremely close to zero.  Such a constraint is easy to achieve, since all it requires is that we fix one fermion mass.  Unfortunately, due to the fact that
$(\Lambda / a)$ represents the cutoff after inflation the usual argument that curvature terms can be inflated away can't be used to eliminate this term.

The remaining terms are either the differences of boson masses
to the fourth power or fermion masses to the fourth power, times logarithms in the cutoff.
Since these terms are independent of $a$ they can represent a cosmological constant.
Now, however, due to the presence of terms involving logs of the cutoff,
choosing appropriate masses can only eliminate them at a single time.
After that, as we already noted, they will change with time with the logarithm of the
scale factor of the universe.  Generically, this means that the FRW equation will look
like
\be
    \left( \dot{a} \over a \right)^2 = { 8 \pi G \over 3} \left(
    {\rho_\gamma \over a^4} + {\rho_m \over a^3} + D + F \ln(a)
    \right) .
\label{FRWequation}
\ee
This argument says that for an appropriately restricted set of free fields there is no
problem achieving a universe that satisfies FRW cosmology with a time dependent ``cosmological
constant''.  In the next section I will argue that since the generic structure of this formula
follows from dimensional analysis, a similar argument can be made for at least some
class of interacting theories that have a fixed number of
quantum degrees of freedom per co-moving volume.  Thus, it behooves us to take a few moments
to extract one phenomenologically significant result from this generic form: namely,
a formula relating the difference of the parameter $w$ from minus one and its time derivative.
For the case of FRW cosmology it is easy to get a formula for $w$ given a formula for the
vacuum energy density as a function of $a$.  The simplest way to do this, that avoids the
Einstein equations, is to observe the the vacuum energy, $E_{vac}$ is given by
\be
E = a^3\,{\cal E}
\ee
and to observe that the definition of pressure says that
\be
 - p dV = {d E \over dV} dV
\ee
Thus, since $dV = 3\,a^2 da$ we have
\be
    -3\,p\,a^2\,da = 3\,a^2\,{\cal E} \,da + a^3 {d{\cal E} \over da} \, da .
\ee
Defining $ p = w {\cal E}$ we immediately obtain
\bea
    - \left( 1 + w \right) &=& {a \over 3 {\cal E}}\,{d{\cal E} \over da} \\
    w &=& -1 - {a \over 3 {\cal E}} {d {\cal E}\over da} .
\eea
Substituting in the contribution to the energy density of the universe
coming from the constant and log terms in our formula we get
\be
   w = -1 -{1 \over 3} { F \over D + F\,\ln(a)} .
\label{w}
\ee
Differentiating with respect to time yields
\be
    \dot{w} = {1 \over 3} \left( { F \over D + F\,\ln(a)} \right)^2\,H(a) .
\label{wdot}
\ee
Thus, if we define the scale factor as being unity at the present time, then we know
that, since experiments show that the ``cosmological constant'' is positive, it follows that
$D > 0$.  While in principle the formula we have could yield a value for $w$ that is less than
$-1$, let us for now assume that $w  > -1$, which tells us that $F < 0$.  From this it follows
that the cosmological constant is decreasing with time, and therefore it was larger in
the past.  The same equations can be used to get more information.  From the formula for $w$
\be
   w = -1 +{1 \over 3} {\vert F \vert \over D - \vert F \vert\,\ln(a)} 
\ee
we see that, for $a = 1$, the difference between $w$ and minus one is given by $ \vert F \vert /D$.
Given current limits on $w$\cite{Hicken}\cite{Frieman} we see that the upper limit on
the error is $ \vert F \vert /D \sim 0.1$.
Thus, we see from Eq.\ref{wdot}, that in order to see a one percent change in $w$ we will have to
wait a Hubble time.  From this we see that one shouldn't be able to measure $\dot{w}$ with
experiments limited in range to small values of $z$.

\section{Taking Interactions into account}

Having warmed up on the free field case, the question is what happens when we consider interacting fields?  Here, of course, the situation becomes less clear.  I already noted in the previous section that the formula I wrote down for the dependence of things on the co-moving cutoff is essentially dimensional analysis.  The question is under what conditions should we expect the required cancellations in the $(\Lambda/a)^4$ and $(\Lambda/a)^2$ terms to persist when interactions are turned on.  While I don't know the most general situation that can be tolerated, it is clear that for unbroken supersymmetry, where the vacuum energy is zero, these terms are not there even when all supersymmetric masses and interactions are turned on.  If we now consider softly broken supersymmetric theories it is plausible that the terms that scale with the cutoff over $a$ do not reappear.  I say it is plausible because supersymmetry doesn't really like a cutoff like this, so one would have to have a lattice theory with a restricted form of lattice supersymmetry that guarantees these two conditions.  Supersymmetric lattice theories have been discussed in the literature; however I don't know if they will do the job\cite{Kaplan}.

For the moment, let us assume that this is possible.  In that event, we once again see that the vacuum energy density will be of the form $ D + F\,\ln(a)$.  Now, since this is an interacting theory it is no longer simple to guarantee that one can arrange the terms that appear in $D$ and $F$ to be small.  However, even if it is a bit ugly, one can always make these terms small at one time by adding a set of free ``shadow fields'' that couple only to gravity and that satisfy the constraints of the previous section.  We can then use these fields to generate counter-terms for the ``time dependent cosmological constant'' and tune the coefficients of the constant and $\ln(a)$ piece.  With this trick it should be possible to create an interacting theory where the general form of the behavior of the vacuum energy density has the same form as in the free case and the same phenomenological limits on $\dot{w}$ should apply.

Obviously, without a lot more work, I can't make more detailed arguments about the
interacting case.  Clearly, the argument that the $(\Lambda/a)^4$ term cancels for
appropriate numbers of fermions and bosons can be expected to be the most robust, at least
so long as $\Lambda/a$ isn't at present too far below the Planck scale.  The curvature term
is more problematic since it would contribute to the percentage of critical density due to
curvature contributions
\be
   \Omega_K \sim { 8\pi \over 3} \left({\Lambda \over M_p} \right)^2
   {\Delta(m_i^2, \mu_i^2) \over H^2(a) a^2} .
\ee
Given the very small value of the Hubble constant at present this implies that the
$\Delta$, which has the dimensions of mass squared, is extraordinarily small.  There is
essentially a factor of $10^{80}/m_p^2$ (where $m_p$ stands for the proton mass)
that needs to be cancelled out.  If, however, this
can be controlled by a symmetry the remaining terms are of the form that can be
renormalized by the spectator fields.

Having made these arguments it is important to ask about other problems that can arise in
an interacting theory.  The key observation in this regard is that in a world where the
effective cutoff is falling as $a$ increases, most physical coupling constants
such as $\alpha_{em}$ will vary with $a$.  This is because counter-terms can only
be used to set the values of these coupling constants and masses for a single value
of $a$.  After that, as $a$ increases, the necessary cancellations stop working and
the couplings run.   Thus, one must ask what bounds exist upon the variation of the
parameters of physics as we know it due to the expansion of the universe.

Probably, the most stringent bound would be upon the behavior of hadron masses
due to changes of the cutoff in QCD.  Fortunately, given the scenario we are
considering, there is no reason to expect this to be the case.  The argument for this
is simple.  We already know from lattice QCD that one gets good answers for ratios of
hadron masses once the lattice spacing is chosen to be small enough and the coupling
is chosen small enough.  After that refining the lattice and moving the coupling to a smaller
value doesn't change these ratios.  Since it is possible to think about the co-moving
cutoff as a theory where a single degree of freedom, the scale factor $a$, is
coupled to a lattice theory (e.g., in Eq.\ref{ffhamiltonian} replace the continuous
free field Hamiltonian with a lattice version), then it is plausible to assume that
so long as $a$ is not so large that $\Lambda /a \ll M_p$, the properties of hadronic
masses and couplings won't be changed.  Thus, we are left thinking about the phenomenological
constraints on time variations of the weak and electromagnetic theories.  Within this
framework the strongest constraints are clearly on variations in the fine structure constant,
$\alpha_{em}$ and the mass of the electron, $m_e$.  The question facing us is ``How do we expect
these quantities to vary as $a$ increases?''.

The answer clearly depends in detail upon the theory, but generically, at the one loop level. it is given by perturbation theory factors times a term of the form $\ln (\Lambda/m\,a)$.  Since we are subtracting away this correction for the current value of $a$ so as to set $\alpha_{em}$ to its current value, we see that as $a$ changes the cancellation that we have introduced is not exact, and we will have a term that goes as the coefficient of the loop amplitude times $\ln( a )$.  To estimate the change in $\alpha_{em}$ from its present value we need to consider data in which the change in $a$ is as large as possible.

Clearly the first place to look is at possible changes in $\alpha_{em}$ from the time of big bang nucleosynthesis (BBN), since between now and then $a$ has changed by a factor of $10^{10}$.  Current limits\cite{Uzan} constrain the variation of $\Delta \alpha_{em} / \alpha_{em}$ to be less than $0.1$.  Given a change in the logarithm of order 10 this means that BBN limits are inconsistent with the idea of a co-moving cutoff at the one loop level.  Thus, in order to evade this constraint one would have to cancel variations in $\alpha_{em}$ and $m_e$ to one loop, which implies constraints among masses and charges of particles in the theory.  Clearly, one can
construct theories where the cancellation occurs, and if one does so,
then these changes would be on the order of $10^{-3}$ and consistent with BBN.

The next place to look for variation of $\alpha_{em}$ in measurements looking at high quality
absorption spectra for QSOs\cite{Webb}\cite{Yang}.  There is a checkered history for these
results and rather than take the statements that something has been seen at a level of $10^{-5}$,
I would rather state that certainly nothing has been seen at a level of $10^{-4}$.  The values
of $z$ are rather small and the logarithm of the change in $a$ for all of this region is
of order unity.  Thus, these results are more or less consistent with the limits placed upon
the variation in $\alpha_{em}$ due to BBN.

Actually, the issue with the Lyman-$\alpha$ forest measurements is more subtle than I have
said.  The reason for this has to do with the fact that I have been estimating the changes in 
constants assuming, as is the case in FRW cosmology, that the scale factor is changing in a spatially uniform manner.  However, it is important to realize that once a region becomes
gravitationally bound, then it is general relativity that matters, not FRW cosmology.  The
fact is that, in the bound region, the expansion has stopped; i.e., the region has 
decoupled from the expansion and the local co-moving volume is no longer growing in size.
Thus, in such a region, constants no longer run.  This means that one could see a time
variation in $\alpha_{em}$  and $m_e$ in the Lyman-$\alpha$ forest measurements, but it
reflects the fact that various regions stopped expanding at different times.  In particular 
this means that the observed values for the constants could show both temporal and 
spatial variation.

Finally, one might worry that the real tension with this idea comes from geological constraints, in particular constraints coming from studies of the Oklo phenomenon.  This analysis corresponds to measurements at a $z \sim 0.1$ and so, suppressing 1-loop effects, as for BBN, we naively expect an  effect on the order of $10^{-5}$.  The Oklo bound is that there has been no change in $\alpha_{em}$ in this period to less than a part in $10^{-8}$\cite{Uzan}.  Similar bounds come from other geological measurements involving $\alpha$-particle decay, etc.  Thus, it would seem that we are faced with the necessity of looking for theories which evade changes in $\alpha_{em}$ to the two-loop level; a situation that becomes very unattractive.   While I could argue that
strong assumptions go into these arguments, in particular that only $\alpha_{em}$  varies, it isn't necessary to worry about this for the reason I just stated above.  Certainly, in order for
the Oklo phenomenon to exist, the Earth and its local surroundings must have already decoupled
from the cosmological expansion of the universe.  Thus, given the assumption that it is the
number of degrees of freedom per co-moving volume that sets the cutoff, there is no reason
for $\alpha_{em}$ to have changed locally over the history of the solar system and all geological bounds are irrelevant.

It should be clear from this discussion that the bounds obtained from FRW considerations are
at most upper bounds on the amount of variation that one might expect to measure.
There is another point to be raised when discussing these phenomenological constraints that
is also related to the way in which I am estimating these changes.  My philosophy on this has been to say that by the time of BBN the theory has settled down to the standard model and to think of everything in terms of an effective theory where constants have been given particular values by adding the necessary counterterms to the effective Lagrangian and then asking how things change with time.  It is not at all clear that one could not have a class of theories for which
the expected changes could be very small.  Suppose, for example, the idea that the true theory is asymptotically conformal are true, then far above the unification scale couplings will run extremely slowly with changes in the scale factor.  In that case it might be that so long as the effective cutoff doesn't go far below the Planck scale the effects I am talking about could be further suppressed .  It is certainly clear that for the phenomenological limits under discussion a better analysis of what happens in specific models is needed.

\section{The Why Now Problem}

There is an amusing possibility that is inherent in this picture.  Since couplings change as a function of the scale factor, it is possible that the expansion of the universe can cause phase transitions.  Leaving out discussion of how couplings change as a function of $a$, the notion that phase transitions can be caused by varying $a$ is actually quite intuitive.  If one takes the picture that what one has is lattice theory where changing $a$ is effectively changing the lattice spacing then any self respecting condensed matter theorist would assume that this could cause the theory to cross any number of quantum critical points.  Thus, we see that the scale factor, a
quantity that is usually thought not to be measurable, now becomes a clock for particle physics. Since the ``Why now?'' problem asks why we live at a time when the matter density and cosmological constant terms are roughly equal in size, there is now the possibility that this question can be rephrased as ``For what theories will the energy density in mass and radiation be equal to that of the cosmological constant when the scale factor (or the lattice spacing) has a specific size?''.
Constructing such theories might be difficult and could well result in Byzantine constructs that only a mother could love; nevertheless, the search for such theories would have the advantage
that it would not require anthropic reasoning.  I am not sure having an ugly theory that
answers the question is worse than arguing over whether one should assume equal probability in $t$ or $\ln (t)$.

\section{Special Relativity}

Finally, it is clear that any introduction of a momentum cutoff is inconsistent with special relativity.  In particular, if as I suggested one thinks of this cutoff as due to a lattice theory coupled to the scale factor, then without special tricks one would expect that the energy momentum relationship for massless particles would look like
\be
    E(k) = \sqrt{ k^2 + c_1\,k^2\,\left(k\,a \over\Lambda\right)^2+\cdots}
\ee
This, of course leads to a change in the speed of light as a function of energy, and so
one could look for signs of variation of the speed of light with energy by looking
at differences in arrival times of photons of different sources from distant
astronomical events or for changes in the relativistic energy momentum dispersion relation
at accelerators.  If one is looking for effects based upon the existence of a lattice,
then unlike arguments about the effects of quantum gravity, the change
in velocity as a function of momentum is quadratic in the ratio of the energy
to the cutoff, $\Lambda$; hence, limits such as those given by the Fermi
collaboration\cite{fermi} become insignificant. From a theoretical point of view
the trouble with trying to put strong limits on the assumed shortest distance using
these experiments is that small changes in the lattice
action can easily suppress these effects to a very high degree, making definitive statements
about the existence of a shortest distance scale difficult to check.  This question is,
however, extremely important since we do not know from these arguments what the lowest
possible value for the effective cutoff, $\Lambda/a$, can be at the present time.

\section{Summary}

In the preceding sections of this paper I have attempted to argue that, surprisingly,
positing the existence of a fundamental momentum cutoff, or existence of a
fundamental shortest distance, that varies with the scale factor of the universe
is less constrained by current phenomenology than one might guess.
I also tried to point out that this assumption leads to an attractive framework for
discussing the existence and possible time dependence of the cosmological constant
and the possible time-variation of other fundamental constants in particle physics.
Probably the nicest feature of this idea is that by using essentially dimensional
analysis one can relate the cosmological constant, $w$ and $\dot{w}$ and show that
given current estimates of the difference of $w$ from $-1$ it will be impossible
for current experiments to measure $\dot{w}$.  Hence, this idea has the immediate plus
that if the new experiments meant to measure $\dot{w}$ see something, the idea of this
sort of a behavior of the cutoff is ruled out.

The second thread of this discussion revolved about the fact that assuming this
kind of cutoff leads one inescapably to the idea that the weak coupling constant,
the fine-structure constant, the ratio of the electron to proton mass, etc. should
all run in a predictable way within any specific model.  It is worth emphasizing
that previous attempts to discuss these ideas always took place in a context
that lacks any sort of specificity; at least that can't be said about this
idea.  I further argued that without studying things in detail,
this idea implies constraints on models that could be consistent
with such a cutoff.  Once again, all but one of these constraints don't
seem too formidable and it would seem that it should be possible to create models
that avoid most of them.  Clearly my discussion of this question is woefully inadequate
and a more serious analysis of these issues within the context of specific models
is required in order to better assess the viability of the idea.
Finally, I wanted to again emphasize the point about the fact that there is something very attractive about having the change in the size of the universe drive phase transitions in the underlying particle physics and therefore opening up the possibility of talking about ``Why now?'' problems in a different way.  Certainly at the moment this is no more than a blue-sky possibility since one may never find an attractive theory that does anything interesting.  Nevertheless, given the alternative it seems worth thinking about.  It is certainly amusing that this idea connects physics in the very large and very small together in a way that never happens in a more conventional approach.  Usually, we put in the particle physics and let it drive the cosmology.  With this point of view cosmology returns the favor and drives the details of the particle physics.

\section{Acknowledgement}

I would like to thank Prof. Alfred Goldhaber for reminding me that for gravitationally bound systems the co-moving volume is no longer changing.  This observation is crucial when 
discussing the bounds on the time variation of fundamental constants, since it removes
geological bounds from consideration.


\end{document}